\documentclass
[superscriptaddress,secnumarabic,amssymb,amsmath,nobibnotes,aps,prd,showkeys,showpacs,nofootinbib,onecolumn,notitlepage,12pt]{revtex4}%
\usepackage{setspace}
\usepackage{color}
\usepackage{amsmath}
\usepackage{amsfonts}
\usepackage{verbatim}
\usepackage{amsmath}
\usepackage{amssymb}
\usepackage{graphicx}
\usepackage[caption=false]{subfig}%
\providecommand{\U}[1]{\protect\rule{.1in}{.1in}}

\newcommand{\be}{\begin{equation}}
\newcommand{\ee}{\end{equation}}

\newcommand{\mincir}{\raise
-3.truept\hbox{\rlap{\hbox{$\sim$}}\raise4.truept\hbox{$<$}\ }}
\newcommand{\magcir}{\raise
-3.truept\hbox{\rlap{\hbox{$\sim$}}\raise4.truept\hbox{$>$}\ }}

\ifx\pdfoutput\relax\let\pdfoutput=\undefined\fi
\newcount\msipdfoutput
\ifx\pdfoutput\undefined\else
\ifcase\pdfoutput\else
\ifx\paperwidth\undefined\else
\ifdim\paperheight=0pt\relax\else\pdfpageheight\paperheight\fi
\ifdim\paperwidth=0pt\relax\else\pdfpagewidth\paperwidth\fi
\fi\fi\fi
\begin{document}
\title{New inflationary exact solution from Lie symmetries}
\author{Andronikos Paliathanasis}
\email{anpaliat@phys.uoa.gr}
\affiliation{Institute of Systems Science, Durban University of Technology, Durban 4000,
South Africa}
\affiliation{Instituto de Ciencias F\'{\i}sicas y Matem\'{a}ticas, Universidad Austral de
Chile, Valdivia 5090000, Chile}

\begin{abstract}
For the inflaton field we determine a new exact solution by using the Lie
symmetry analysis. Specifically, we construct a second-order differential
master equation for arbitrary scalar field potential by assuming that the
spectral index for the density perturbations $n_{s}$ and the scalar to tensor
ratio $r$ are related as $n_{s}-1=h\left(  r\right)  $. Function $h\left(
r\right)  $ is classified according to the admitted Lie symmetries for the
master equation. The possible admitted Lie symmetries form the $A_{2}$,
$A_{3,2}$, $A_{3,3}$ and $sl\left(  3,R\right)  $ Lie algebras. The new
inflationary solution is recovered by the Lie symmetries of the $A_{3,3}$
algebra. Scalar field potential is derived explicitly, while we compare the
resulting spectral indices with the observations.

\end{abstract}
\maketitle
\date{\today}

\section{Introduction}

\label{sec1}

Inflation is the main mechanism to explain the homogeneity and isotropy of the
observable universe at present time. During the inflationary period, the
universe was dominated by the inflaton. The inflaton drove the dynamics such
that to provide the expansion \cite{guth}. Nevertheless, the inflationary
models are mainly defined on homogeneous spacetimes, or on background spaces
with small inhomogeneities \cite{st1,st2}.

In \cite{w1} it was found that the presence of a positive cosmological
constant in Bianchi cosmologies leads to expanding Bianchi spacetimes,
evolving toward an rapidly expanding de Sitter universe. That was the first
result to support the cosmic \textquotedblleft no-hair\textquotedblright%
\ conjecture \cite{nh1,nh2}. The conjecture states that all expanding
universes with a positive cosmological constant admit as asymptotic expanding
solution. The necessity of the rapid expansion is that it provides a rapid
expansion for the size of the universe such that the latter effectively loses
its memory on the initial conditions, which means that the rapid expansion
solves the \textquotedblleft flatness\textquotedblright, \textquotedblleft
horizon\textquotedblright\ and monopole problem \cite{f1,f2}.

Because inflation occurred in a finite time period during the cosmological
evolution, instead of the cosmological constant, the inflaton is assumed to be
described by a scalar field, which provides a dynamical behavior for the
evolution of the field equations. In the single scalar field inflationary
models the expansion appears when the scalar field potential dominates. In
these models, the inflationary period is determined explicitly by the nature
of the scalar field potential. In the literature, there is a plethora of
scalar field potentials which have been proposed the last decades, see for
instance
\cite{ref1a,ref1,ref2,ref3,newinf,ref4,ref5a,ref5,ref6a,ref6,ref7,ref8,ref9,ref10,ref11}%
.

Moreover, single scalar field gravitational models attribute the additional
degrees of freedom provided by modified theories of gravity, such in the
quadratic gravity \cite{sa1}. The quadratic gravity is an geometric approach
for the construction of the scalar field potential, through a conformal
transformation with the use of a Lagrange multiplier \cite{sot1}. The
quadratic gravity, is also known as Starobinsky model for inflation and it is
the model which is mainly supported by the recent cosmological observations
\cite{pin1}. There is a plethora of inflationary models which have been
proposed the last years, multifield inflationary models or inflationary models
which follow from modifications of the Einstein-Hilbert Action Integral, see
for instance
\cite{Aref2,Aref3,Aref3b,Aref5,Aref6,Aref6b,Aref7,ar1,ar2,ar3,ar4}. A
catalogue with the viable inflationary models published in \cite{ec1}, while
the updated version published later in \cite{ec2} presents the viable
inflationary models after the release of Planck 2013 data.

The field equations for the inflaton model are that of a minimally coupled
scalar field in the context of General Relativity with a spatially flat
Friedmann--Lema\^{\i}tre--Robertson--Walker (FLRW) as a background space. The
field equations are nonlinear differential equations of second-order and form
a singular Lagrangian system with dynamical variables the scalar factor of the
metric tensor. The existence of exact solutions for the field equations
depends on the functional form of the scalar field potential \cite{ell1,ell2}
which drives the scalar field dynamics. In \cite{sym1,sym2}, the scalar field
potential has been constraint such that the field equations to admit
conservation laws. The existence of the conservation laws indicates the
Liouville integrability for the field equations such that exact and analytic
solutions to be determined. However, an innovative approach for the
determination of analytic solutions in scalar field cosmology applied in
\cite{ns1}. Because the field equations form a singular dynamical system,
there exist infinity number of nonlocal conservation laws, which can be used
to reduce the order for the field equations. Indeed, with the use of the
nonlocal conservation laws, the generic algebraic solution has been found, for
arbitrary potential function. The results of \cite{ns1} applied for the
derivation of new inflationary solutions in \cite{ref10}, while this approach
applied for the reconstruction of the inflationary potential in \cite{ref11}.

Specifically, in \cite{ref11} it was assumed that spectral index for the
density perturbations, $n_{s}$, and the tensor-to-scalar ratio, $r$, are
related by a function such that $n_{s}-1=h\left(  r\right)  $. where $h\left(
r\right)  $ is either constant, linear or quadratic function in $r$. By
applying the results of \cite{ns1} in each case a master differential equation
has been defined which gives the resulting inflationary potential. The
resulting inflationary solutions fits the cosmological observations, while the
closed-form expressions for the scalar field potential were found. In this
piece of work, we focus on the classification of function $h\left(  r\right)
$ according to the admitted Lie symmetries for the master equations. This
geometric selection rule is inspired by the results of the applications of the
symmetry analysis in \cite{sym1}. The classification scheme that we follow, it
was established by Ovsiannikov on the classification of the unknown function
for a nonlinear Schr\"{o}dinger equation \cite{ovs1}. A similar reconstruction
approach for the slow-roll inflation can be found in the series of studies
\cite{ff1,ff2} where the scalar-to-tensor ratio has been assumed to be a
specific function with independent variable the number of e-folds. The
corresponding scalar-tensor theory and the resulting $f\left(  R\right)
$-theory determined. The plan of the paper is as follows.

In Section \ref{sec2} we define the gravitational model for a single scalar
field inflationary theory and we present Einstein's field equations. Moreover,
we present the generic algebraic solution for arbitrary potential found in
\cite{ns1}, and we define the master equation of our analysis. The Lie
symmetry analysis for the master equation of our consideration is performed in
Section \ref{sec3}. In Section \ref{sec4} we present the new inflationary
exact solution as also we calculate the spectral indices $n_{s}$ and $r$.
Finally, in Section \ref{sec6} we draw our conclusions.

\section{The inflaton field}

\label{sec2}

Consider the Action Integral of Einstein-Hilbert Action with a scalar field
minimally coupled to gravity, that is,%
\begin{equation}
S=\int dx^{4}\sqrt{-g}\left[  R-\frac{1}{2}g^{\mu\nu}\phi_{;\mu}\phi_{;\nu
}+V(\phi)\right]
\end{equation}
in which $R$ is the Ricci-scalar for the background space with metric tensor
$g_{\mu\nu}\left(  x^{\kappa}\right)  $, $\phi\left(  x^{\kappa}\right)  $ is
the scalar field and $V\left(  \phi\left(  x^{\kappa}\right)  \right)  $ is
the scalar field potential. Variation with respect to the metric tensor
provides the Einstein-field equations%
\begin{equation}
R_{\mu\nu}-\frac{1}{2}g_{\mu\nu}R=T_{\mu\nu}^{\left(  \phi\right)  }
\label{sfa.01}%
\end{equation}
where\ $R_{\mu\nu}$ is the Ricci tensor,\ and $T_{\mu\nu}^{(\phi)}$ is the
energy-momentum tensor which corresponds to the scalar field. $T_{\mu\nu
}^{\left(  \phi\right)  }$ is defined as%
\begin{equation}
T_{\mu\nu}^{\left(  \phi\right)  }=\phi_{;\mu}\phi_{;\nu}-g_{\mu\nu}\left(
\frac{1}{2}\phi^{;\kappa}\phi_{;\kappa}-V\left(  \phi\right)  \right)  .
\end{equation}

Furthermore, variation with respect to the scalar field in the Action
Integral, provides the equation of motion for the scalar field, that is, the
Klein-Gordon equation $T_{~~~~~~~~~;\nu}^{\left(  \phi\right)  \mu\nu}=0,$
that is,
\begin{equation}
g^{\mu\nu}\phi_{;\mu\nu}-V\left(  \phi\right)  =0.
\end{equation}

For the background space we consider a spatially flat FLRW universe with line
element%
\begin{equation}
ds^{2}=-dt^{2}+a^{2}\left(  t\right)  \left(  dx^{2}+dy^{2}+dz^{2}\right)  .
\end{equation}
Function $a\left(  t\right)  $ is the scale factor while the Hubble function
is defined as $H=\frac{\dot{a}\,}{a}$,~an overdot denote differentials with
respect to comoving proper time, $t$.

Hence, the Einstein's field equations (\ref{sfa.01}) are%
\begin{equation}
3H^{2}=\frac{1}{2}\dot{\phi}^{2}+V(\phi), \label{in.03}%
\end{equation}%
\begin{equation}
2\dot{H}+3H^{2}=-\frac{1}{2}\dot{\phi}^{2}+V(\phi), \label{in.04}%
\end{equation}
while the Klein-Gordon equation reads
\begin{equation}
\ddot{\phi}+3H\dot{\phi}+V_{,\phi}=0. \label{in.05}%
\end{equation}
We have assumed that field $\phi$ inherits the symmetries of the background
space, that is, $\phi=\phi\left(  t\right)  $. Inflation occurs, when the
scalar field potential dominates, $3H^{2}\simeq V\left(  \phi\right)  $ and
the scalar field approaches a stationary point $\dot{\phi}\simeq
-\frac{V_{,\phi}}{3H}~$\cite{l01}.

Thus in order to understand the existence of an inflationary era provided by a
potential function, the potential slow-roll parameters (PSR)
\begin{equation}
\varepsilon_{V}=\left(  \frac{V_{,\phi}}{2V}\right)  ^{2}~\,,~\eta_{V}%
=\frac{V_{,\phi\phi}}{2V}, \label{in.08}%
\end{equation}
have been introduced \cite{l01}.

The condition an inflationary era to exists is expressed as, $\varepsilon
_{V}<<1$, while in order for the inflationary phase to last long enough it is
required require the second PSR parameter also to be small, i.e. $\eta_{V}<<1$.

Some inflationary potentials which have been proposed in the literature are
presented. The quadratic $R+\left(  \frac{R}{6M}\right)  ^{2}$ inflationary
model with potential function in the Einstein-frame \cite{sa1}
\begin{equation}
V_{A}\left(  \phi\right)  =V_{0}\left(  1-e^{-\sqrt{\frac{2}{3}}\frac{\phi}%
{M}}\right)  ^{2},
\end{equation}
the intermediate inflationary potential \cite{oll0}%
\begin{equation}
V_{B}\left(  \phi\right)  =\frac{48A^{2}}{\left(  \Delta+4\right)  ^{2}%
}\left(  \frac{\phi-\phi_{0}}{\sqrt{2A\Delta}}\right)  ^{-\Delta}%
-\frac{4A\Delta}{\left(  \Delta+4\right)  ^{2}}\left(  \frac{\phi-\phi_{0}%
}{\sqrt{2A\Delta}}\right)  ^{-\Delta-2}~,
\end{equation}
the natural inflation \cite{oll1}%
\begin{equation}
V_{C}\left(  \phi\right)  =V_{0}\left(  1+\cos\left(  \frac{\phi}{f}\right)
\right)  ,
\end{equation}
the hyperbolic inflationary potential \cite{ref7}%
\begin{equation}
V_{D}\left(  \phi\right)  =V_{0}\sinh^{q}\left(  p\right)  ~,
\end{equation}
and many others. For an extended list of the proposed inflationary models we
refer the reader in the reviews \cite{ec1,ec2}.

Similarly, with the PSR parameters we can define the corresponding Hubble
slow-roll parameters (HSR) as \cite{l01}\qquad%
\begin{equation}
\varepsilon_{H}=-\frac{d\ln H}{d\ln a}=\left(  \frac{H_{,\phi}}{H}\right)
^{2}, \label{in.09}%
\end{equation}%
\begin{equation}
\eta_{H}=-\frac{d\ln H_{,\phi}}{d\ln a}=\frac{H_{,\phi\phi}}{H}. \label{in.10}%
\end{equation}

The two different set of parameters, the HSR and PSR parameters are related by
the expressions
\begin{equation}
\varepsilon_{V}\simeq\varepsilon_{H}~\text{and~}\eta_{V}\simeq\varepsilon
_{H}+\eta_{H}. \label{in.11}%
\end{equation}
Therefore, inflation occurs when $\varepsilon_{H}<<1$. The limit
$\varepsilon_{H}=1$, is called the end of inflation where we have the exit
from the acceleration era. In the following we work with the Hubble slow-roll parameters.

As far as the observable values for the spectral indices $n_{s}$ and $r$, are
concerned, from the Planck 2018 collaboration \cite{pin2018} follow that the
spectral index for the density perturbations is constraint as $n_{s}%
=0.9649\pm0.0042,$ while the tensor to scalar ratio, $r$ is constraint as
$r<0.10$.

These indices are related with the HSR parameters in the first-order
approximation as%
\begin{equation}
r=10\varepsilon_{H}~,
\end{equation}%
\[
n_{s}=1-4\varepsilon_{H}+2\eta_{H}.
\]

In the second-order approximation we shall consider the additional slow-roll
parameter $\xi_{H}\equiv\frac{H_{\phi}H_{\phi\phi\phi}}{H^{2}}$, such that
$n_{s}$ and $n_{s}^{\prime}$ are expressed as
\begin{equation}
n_{s}\equiv1-4\varepsilon_{H}+2\varepsilon_{H}-8\left(  \varepsilon
_{H}\right)  ^{2}\left(  1+2C\right)  +\varepsilon_{H}\eta_{H}\left(
10C+6\right)  -2C\xi_{H},
\end{equation}%
\begin{equation}
n_{s}^{\prime}\equiv2\varepsilon_{H}\eta_{H}-2\xi_{H}.
\end{equation}
where $C=\gamma_{E}+\ln2-2\simeq-0.7296$ and the range of the scalar spectral
index is $n_{s}^{\prime}=-0.005\pm0.013$ \cite{pin2018}.

In the following we consider the spectral indices in the first-order approximation.

\subsection{Algebraic solution}

In \cite{ns1}, the generic algebraic solution for the field equations
(\ref{in.03})-(\ref{in.05}) is presented for arbitrary potential function.
Indeed, with the use of new variables the unknown potential function can be
introduced inside the metric tensor, while the dynamical variables for the
field equations are expressed in terms of an arbitrary function. The field
equations are reduced in algebraic equation. Such a solution is called
algebraic solution.

We define the new variable $dt=\exp\left(  \frac{F\left(  \omega\right)  }%
{2}\right)  d\omega$ with $\omega=6\ln a$. The FLRW line element becomes
\begin{equation}
ds^{2}=-e^{F\left(  \omega\right)  }d\omega^{2}+e^{\omega/3}(dx^{2}%
+dy^{2}+dz^{2}).
\end{equation}
with Hubble function $H\left(  \omega\right)  =\frac{1}{6}e^{-\frac{F}{2}}$.

Hence, from the field equations (\ref{in.03})-(\ref{in.05}) we find that then
the scalar field is expressed as \cite{ns1}%
\begin{equation}
\phi(\omega)=\pm\frac{\sqrt{6}}{6}\int\!\!\sqrt{F^{\prime}(\omega)}d\omega
\end{equation}
while the scalar field potential reads%
\begin{equation}
V(\omega)=\frac{1}{12}e^{-F(\omega)}\left(  1-F^{\prime}(\omega)\right)  .
\end{equation}
in which $F^{\prime}\left(  \omega\right)  =\frac{dF\left(  \omega\right)
}{d\omega}$.

Function $F\left(  \omega\right)  $ can be constraint by various approaches.
In \cite{ref10} closed-form scalar field solutions were found by assuming
specific functional forms for the equation of state parameter $P_{\phi
}=P_{\phi}\left(  \rho_{\phi}\right)  $ of the scalar field, $P_{\phi}%
=\frac{1}{2}\dot{\phi}^{2}-V\left(  \phi\right)  $, $\rho_{\phi}=\frac{1}%
{2}\dot{\phi}^{2}+V(\phi)$.

In the new set of variables, the \ HSR parameters are expressed as follows
\cite{ref10}%
\begin{equation}
\varepsilon_{H}=3F^{\prime}~,~\eta_{H}=3\frac{\left(  F^{\prime}\right)
^{2}-F^{\prime\prime}}{F^{\prime}}, \label{ss01}%
\end{equation}
while the number of e-folds is defined as%
\begin{equation}
N_{e}=\int_{t_{i}}^{t_{f}}H\left(  t\right)  dt=\ln\frac{a_{f}}{a_{i}}%
=\frac{1}{6}\left(  \omega_{f}-\omega_{i}\right)  .
\end{equation}

In \cite{ref10} it was considered that the spectral indices are constraint as
\begin{equation}
n_{s}-1=h\left(  r\right)  , \label{ss1}%
\end{equation}
where $h\left(  r\right)  $ has been assumed to be $h\left(  r\right)
=h_{1}+h_{2}r+h_{3}r^{2}.~$For this latter assumption, new exact and analytic
inflationary solutions were found.

In the first-order approximation, from (\ref{ss1}) and (\ref{ss01}) the master
equation (\ref{ss1}) reads%
\begin{equation}
F^{\prime\prime}+G\left(  F^{\prime}\right)  =0,~G\left(  F^{\prime}\right)
=-\frac{1}{6}h\left(  F^{\prime}\right)  F^{\prime}+\left(  F^{\prime}\right)
^{2}. \label{ss2}%
\end{equation}

In the following we constrain function $G\left(  F^{\prime}\right)  $, i.e.
$h\left(  F^{\prime}\right)  $, by the requirement the master equation
(\ref{ss2}) to admit Lie point symmetries. This selection rule has geometric
characteristics. Symmetries are geometric objects in the space where the
differential equation is defined. Such a geometric selection rule is in
agreement with the geometric nature of gravitational physics, for a discussion
we refer to \cite{sym1}.

We proceed with the presentation of the symmetry analysis for the master
equation (\ref{ss2}).

\section{Lie symmetry analysis}

\label{sec3}

The theory of symmetries of differential equations established at the end of
the 19th century by Sophus Lie \cite{kumei}. The novelty of Lie's approach is
that the transformations group which leave invariant a differential equation
can be used to simplify the given equation. In the case of ordinary
differential equations the Lie point symmetries are used to reduce the order
of the differential equation.

Let us demonstrate this on our master equation (\ref{ss2}). By assuming the
new variables $f=F^{\prime}$, \ the master equation (\ref{ss2}) can be written
in the equivalent form $f^{\prime}=G\left(  f\right)  $, that is $\int
\frac{df}{G}=\omega-\omega_{0}$. This is the most common reduction process
which holds for autonomous dynamical systems where the autonomous point
symmetry vector $\partial_{\omega}$ exists.

The derivation of Lie (point) symmetries for a given dynamical system is based
on a simple algorithm. We briefly discuss the algorithm.

Consider the infinitesimal point transformation
\begin{align}
\bar{\omega}  &  =\omega+\varepsilon\xi\left(  \omega,F\right)  ~,~\\
\bar{F}  &  =F+\varepsilon\eta\left(  \omega,F\right)  ~,
\end{align}
with generator the vector field $X=\xi\left(  \omega,F\right)  \partial
_{\omega}+\eta\left(  \omega,F\right)  \partial_{F}$. Then, the master
equation (\ref{ss2}) is invariant under the Action of the later point
transformation if and only if%
\begin{equation}
X^{\left[  2\right]  }\left(  F^{\prime\prime}-G\left(  F^{\prime}\right)
\right)  =0 \label{ss3}%
\end{equation}
in which $X^{\left[  2\right]  }=X+\eta^{\left[  1\right]  }\partial
_{F^{\prime}}+\eta^{\left[  2\right]  }\partial_{F^{\prime\prime}}$ is the
second extension of $X$ in the tangent space, where $\eta^{\left[  1\right]
}$ and $\eta^{\left[  2\right]  }$ are defined as
\begin{align}
\eta^{\left[  1\right]  }  &  =\eta^{\prime}-F^{\prime}\xi^{\prime},\\
\eta^{\left[  2\right]  }  &  =\eta^{\left[  1\right]  ^{\prime}}%
-F^{\prime\prime}\xi^{\prime}.
\end{align}

When for a given point transformation the symmetry condition (\ref{ss3}) is
valid, the vector field $X$ will be called a Lie symmetry.

The symmetry condition (\ref{ss3}) for the master equation (\ref{ss2}) is
expressed as follow%
\begin{align}
0  &  =\eta_{,\omega\omega}-F^{\prime}\xi_{,\omega\omega}+\left(  F^{\prime
}\right)  ^{2}\left(  \eta_{,FF}-2\xi_{,\omega F}\right)  -\left(  F^{\prime
}\right)  ^{3}\xi_{,FF}+2F^{\prime}\eta_{,\omega F}+G_{,F^{\prime}}%
\eta_{,\omega}+\nonumber\\
&  +\left(  G_{,F^{\prime}}F^{\prime}-G\right)  \eta_{,F}+\left(  3GF^{\prime
}-G_{,F^{\prime}}\left(  F^{\prime}\right)  ^{2}\right)  \xi_{,F}+\left(
2G-G_{,F^{\prime}}F^{\prime}\right)  \xi_{,\omega}. \label{ss4}%
\end{align}

From the later equation we can define systems of partial differential
equations which shall constraint the unknown functions $\xi\left(
\omega,F\right)  $ and $\eta\left(  \omega,F\right)  $. However the constraint
system it depends on the functional form of $G\left(  F^{\prime}\right)  $.
Indeed, for arbitrary function $G\left(  F^{\prime}\right)  $ we find the
generic solution $\xi\left(  \omega,F\right)  =\alpha_{1}$ and $\eta\left(
\omega,F\right)  =\alpha_{2}$, which means that the generic symmetry vector is
$X=\alpha_{1}\partial_{\omega}+\alpha_{2}\partial_{F}$. Hence, the two
independent symmetries, $X_{1}=\partial_{\omega}$ and $X_{2}=\partial_{F}$,
follow, with commutator $\left[  X_{1},X_{2}\right]  =0.$

However, there are special forms of $G\left(  F^{\prime}\right)  $ in which
the master equation (\ref{ss2}) admits additional Lie symmetries. This
classification problem is that we attempt to solve in this study. It is
important to mention that the Lie point symmetries for second-order
differential equations have been widely studied in the literature see for
instance \cite{cs1,cs2}.

It is well known that a second-order differential equation can admits $0,$
$1,~2,~3$ or $8$ Lie symmetries. When eight Lie point symmetries exist, the
the differential equation is called maximal symmetric and there exist a
similarity transformation such that the equation to be written in the form of
the free particle.

For arbitrary function $G\left(  F^{\prime}\right)  $ the master equation
(\ref{ss2}) admits two Lie symmetries, thus we investigate for specific forms
of $G\left(  F^{\prime}\right)  $ in which the master equation admits $3$ or
$8$ Lie symmetries.

Thus, the classification scheme provides for equation (\ref{ss2}) gives the
three functional forms $G_{A}\left(  F^{\prime}\right)  =G_{0}\left(
F^{\prime}\right)  ^{\nu+1},~\nu\neq-1,0,1,2$, $G_{B}\left(  F^{\prime
}\right)  =G_{0}\exp\left(  \mu F^{\prime}\right)  ,~\mu\neq0$ and
$G_{C}\left(  F^{\prime}\right)  =G_{3}\left(  F^{\prime}\right)  ^{3}%
+G_{2}\left(  F^{\prime}\right)  ^{2}+G_{1}F^{\prime}+G_{0}$, where $G_{0-3}$
and $\mu~$are constants parameters

For $G_{A}\left(  F^{\prime}\right)  $, equation (\ref{ss2}) admits the three
Lie symmetries $\left\{  X_{1},X_{2},X_{3}=\nu\omega\partial_{\omega}+\left(
\nu-1\right)  F\partial_{F}\right\}  $ with commutators $\left[  X_{1}%
,X_{2}\right]  =0,~\left[  X_{1},X_{2}\right]  =\nu X_{1}$ and $\left[
X_{2},X_{3}\right]  =\left(  \nu-1\right)  X_{2}$.

For the exponential function $G_{B}\left(  F^{\prime}\right)  $, the Lie
symmetries of equation (\ref{ss2}) are $\left\{  X_{1},X_{2},\bar{X}_{3}%
=\mu\omega\partial_{x}+\left(  F\mu-\omega\right)  \partial_{F}\right\}  $
with commutators $\left[  X_{1},\bar{X}_{3}\right]  =\mu X_{1}-X_{2}$ and
$\left[  X_{2},\bar{X}_{3}\right]  =\mu X_{2}$.

Finally, for $G_{C}\left(  F^{\prime}\right)  $ equation (\ref{ss2}) is
maximal symmetric and admits eight Lie symmetries. The representation of the
admitted Lie algebra depends on the values of the free parameters thus we omit
it. The later case has been widely studied before in \cite{ref11}. Moreover,
the similarity transformations which connect the different inflationary models
which belong to the family of $G_{C}\left(  F^{\prime}\right)  $ are presented
in \cite{sc1}.

For the inflationary models provided by $G_{A}\left(  F^{\prime}\right)  $ and
$G_{B}\left(  F^{\prime}\right)  $ we derive the corresponding functions
$h\left(  r\right)  $ are $h_{A}\left(  r\right)  =-6\left(  h_{0}r^{\nu
}-\frac{1}{10}r\right)  $ and $h_{B}\left(  r\right)  =-6\left(  h_{0}%
r^{-1}e^{\mu r}-\frac{1}{10}r\right)  $.

From the cosmological observations we know that $n_{s}-1\simeq0$, while
$r<0.11$. Thus, from these two models, $h_{A}\left(  r\right)  $ and
$h_{B}\left(  r\right)  $, only model $h_{A}\left(  r\right)  $ provides a
behaviour $\lim_{r\rightarrow0}\left(  h_{A}\right)  \simeq0$ for $\nu>0.~$On
the other hand, $h_{A}\left(  r\right)  <<1$ when $r<<1$, and for $\nu<-1$ if
and only if $h_{0}\simeq\frac{1}{r^{\nu+1}}$, that is, for very large values
of the free parameter $h_{0}$.

In the following section we focus our analysis on model $h_{A}\left(
r\right)  $, where we investigate the closed-form solution, we discuss the
physical properties of the model and we investigate the evolution for the
inflationary parameters.

\section{New inflationary exact solution}

\label{sec4}

For $G=G_{A}\left(  F^{\prime}\right)  $, the master equation reads
\begin{equation}
F^{\prime\prime}+G_{0}\left(  F^{\prime}\right)  ^{\nu+1}=0,
\end{equation}
with closed-form solution
\begin{equation}
F\left(  \omega\right)  =\frac{\left(  G_{0}\nu\omega\right)  ^{1-\frac{1}%
{\nu}}}{G_{0}\left(  \nu-1\right)  }\text{.} \label{f0}%
\end{equation}

Thus, the physical parameters for the inflaton field, such as the energy
density $\rho_{\phi}$, the pressure $P_{\phi}$ and the equation of state
parameter $w_{\phi}=\frac{P_{\phi}}{\rho_{\phi}}$ can be constructed
analytical
\begin{equation}
\rho_{\phi}\left(  \omega\right)  =\frac{1}{12}\exp\left(  \frac{\left(
G_{0}\nu\omega\right)  ^{1-\frac{1}{\nu}}}{G_{0}\left(  \nu-1\right)
}\right)  ~,
\end{equation}%
\begin{equation}
P_{\phi}\left(  \omega\right)  =\frac{1}{12}\exp\left(  \frac{\left(  G_{0}%
\nu\omega\right)  ^{1-\frac{1}{\nu}}}{G_{0}\left(  \nu-1\right)  }\right)
\left(  -1+2\left(  G_{0}\nu\omega\right)  ^{-\frac{1}{\nu}}\right)  ~,
\end{equation}
and%
\begin{equation}
w_{\phi}\left(  \omega\right)  =\left(  -1+2\left(  G_{0}\nu\omega\right)
^{-\frac{1}{\nu}}\right)  .
\end{equation}

We remark that for $\nu>0$ and for large values of $\omega,~w_{\phi}\left(
\omega\right)  \rightarrow-1$, otherwise when $\nu<0$, $w_{\phi}\left(
\omega\right)  \rightarrow-1$ for small values of $\omega$.

As far as the scalar field and the scalar field potential are concerned, we
derive the following expressions%
\begin{equation}
\phi\left(  \omega\right)  =\sqrt{\frac{2}{3}}\frac{\left(  G_{0}\nu
\omega\right)  ^{1-\frac{1}{2\nu}}}{G\left(  2\nu-1\right)  }~,~\nu\neq
\frac{1}{2}, \label{f1}%
\end{equation}%
\begin{equation}
\phi\left(  \omega\right)  =\sqrt{\frac{2}{3}}\frac{\ln\omega}{G_{0}}%
~,~\nu=\frac{1}{2}~, \label{f2}%
\end{equation}
and
\begin{equation}
V\left(  \omega\right)  =\frac{1}{12}\left(  1-\left(  G_{0}\nu\omega\right)
^{-\frac{1}{\nu}}\right)  \exp\left(  \frac{\left(  G_{0}\nu\omega\right)
^{1-\frac{1}{\nu}}}{G_{0}\left(  \nu-1\right)  }\right)  \text{.} \label{f3}%
\end{equation}

In terms of the scalar filed $\phi$, the potential function $V\left(
\phi\right)  $ is given by the following functional form%
\begin{equation}
V\left(  \phi\right)  =\frac{1}{12}\left(  1-\left(  \frac{3}{2}\right)
^{\frac{1}{1-2\nu}}\left(  \left(  2\nu-1\right)  G_{0}\phi\right)  ^{\frac
{2}{1-2\nu}}\right)  \exp\left(  \frac{\left(  \frac{2}{3}\right)
^{\frac{1-\nu}{2\nu-1}}}{G_{0}\left(  \nu-1\right)  }\left(  \left(
2\nu-1\right)  G_{0}\phi\right)  ^{1+\frac{1}{2\nu-1}}\right)  ~,~\nu\neq
\frac{1}{2} \label{f4}%
\end{equation}
or%
\begin{equation}
V\left(  \phi\right)  =\frac{1}{12}\left(  1-\frac{4}{G_{0}}\exp\left(
-\sqrt{6}G_{0}\phi\right)  \right)  \exp\left(  -\frac{4}{G_{0}}\exp\left(
-\sqrt{\frac{3}{2}}G_{0}\phi\right)  \right)  ~,~\nu=\frac{1}{2}\text{.}
\label{f5}%
\end{equation}

\begin{figure}[ptb]
\centering\includegraphics[width=0.6\textwidth]{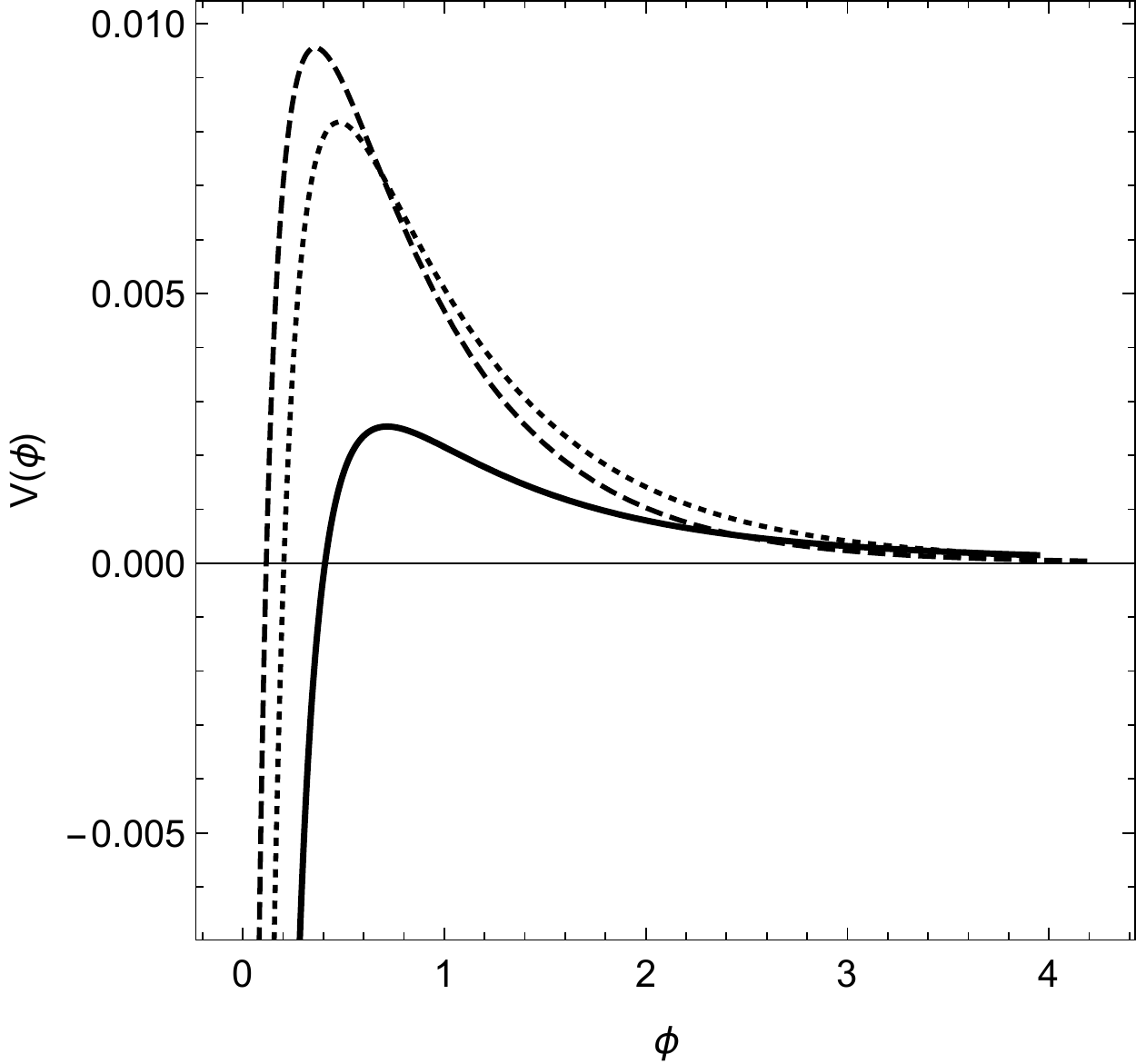}\caption{Qualitative
evolution for the scalar field potential (\ref{f4}) for $G_{0}=1$. Solid line
is for $\nu=\frac{3}{2}$, dotted line is for $\nu=\frac{5}{2}$ and dashed line
is for $\nu=4$. }%
\label{in1}%
\end{figure}

In Fig. \ref{in1} we present the qualitative evolution for the scalar field
potential (\ref{f4}) for $G_{0}=1$ and for different values of the free
parameter $\nu$.

According to our knowledge, this inflationary potential has not been presented
before in the literature.

\subsection{Spectral indices}

For the closed-form solution of $F\left(  \omega\right)  $, the definition of
the slow-roll parameters from expressions (\ref{ss01}) we can define the
spectral indices $n_{s}$ and $r$ in terms of the number of e-fold $N_{e}$.
Recall, that the end of inflation occurs when $\varepsilon\left(  \omega
_{f}\right)  =1$, that is, $\omega_{f}=\frac{3^{\nu}}{G_{0}\nu}$.

Consequently, the spectral indices by using the HSR parameters are written as
follows%
\begin{equation}
n_{s}=1-6\left(  \left(  G_{0}\nu\right)  \left(  6N_{e}+\frac{3^{\nu}}%
{G_{0}\nu}\right)  \right)  ^{-\frac{1}{\nu}}+\frac{6G_{0}}{3^{\nu}%
+6G_{0}N_{e}\nu}~, \label{f6}%
\end{equation}%
\begin{equation}
r=30\left(  \left(  G_{0}\nu\right)  \left(  6N_{e}+\frac{3^{\nu}}{G_{0}\nu
}\right)  \right)  ^{-\frac{1}{\nu}}. \label{f7}%
\end{equation}

We observe that the indices $n_{s}$ and $r$ depend on the number of e-fold
$N_{e}$, and on the free parameters $\left(  \nu,G_{0}\right)  .~$From the
observations the number of e-fold it is considered to be in the range
$N_{e}=\left(  50,60\right)  $.

\begin{figure}[ptb]
\centering\includegraphics[width=1\textwidth]{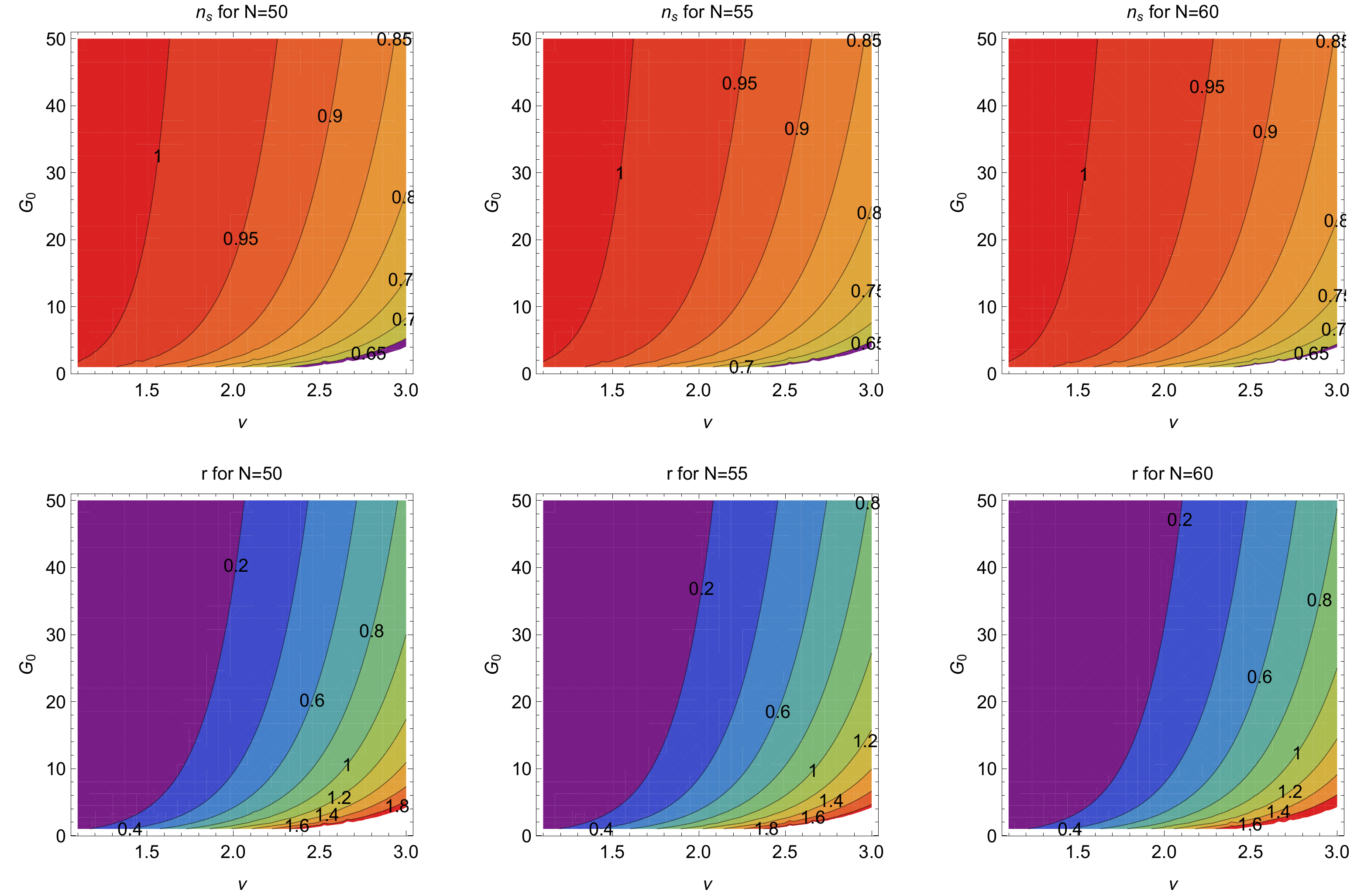}\caption{Qualitative
evolution for the spectral indices $\left(  n_{s},r\right)  $ in the first
approximation in the two-dimensional space for the free variables $\left(
\nu,G_{0}\right)  $, for $N_{e}=50$ (left figs.) $N_{e}=55~$(center figs.) and
$N_{e}=60~$(right figs.)}%
\label{in2}%
\end{figure}

\begin{figure}[ptb]
\centering\includegraphics[width=1\textwidth]{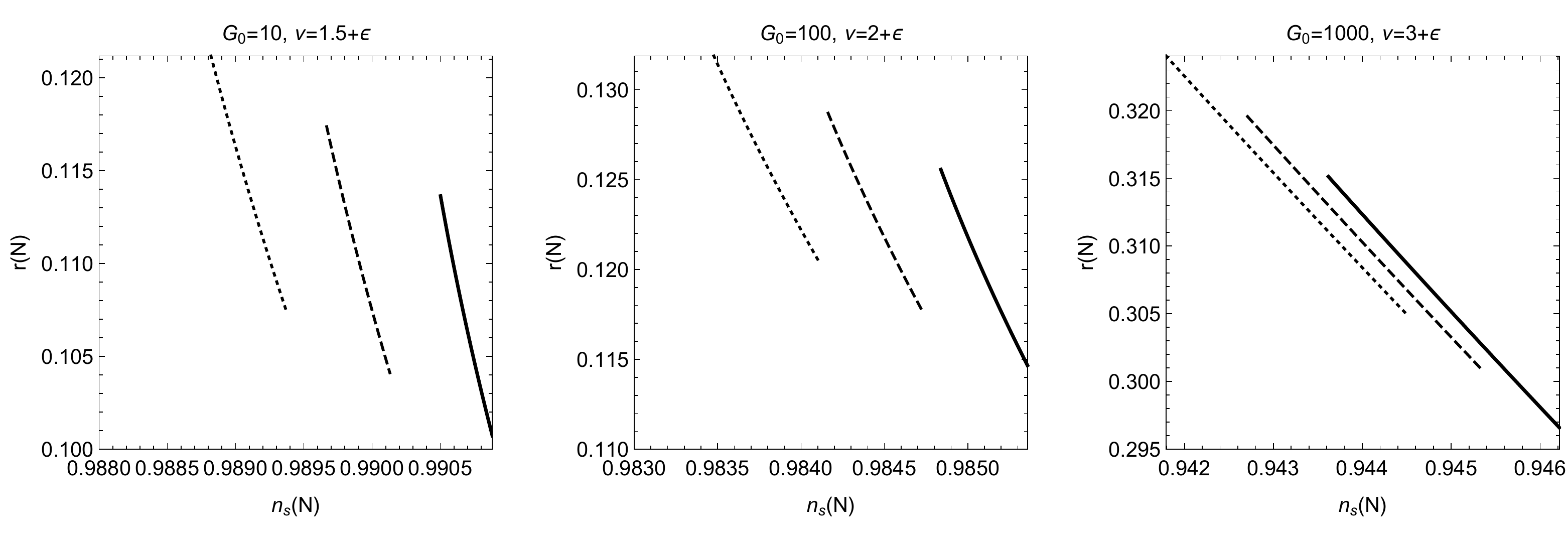}\caption{Parametric
plots for the spectral indices $\left(  n_{s},r\right)  $ in the first
approximation varying number of e-folds $N_{e}=\left(  50,55,60\right)  $ and
different values of $G_{0}$ and $\nu=\nu_{0}+\varepsilon$. Left fig. is for
$\left(  G_{0},\nu_{0}\right)  =\left(  10,1.5\right)  $, center fig. is for
$\left(  G_{0},\nu_{0}\right)  =\left(  10^{2},2\right)  $ and right fig. is
for $\left(  G_{0},\nu_{0}\right)  =\left(  10^{3},3\right)  $. Solid lines
are for $\varepsilon=0.01,~$dashed lines are for $\varepsilon=0.02$ and dotted
lines are for $\varepsilon=0.03$. }%
\label{in3}%
\end{figure}

\begin{figure}[ptb]
\centering\includegraphics[width=1\textwidth]{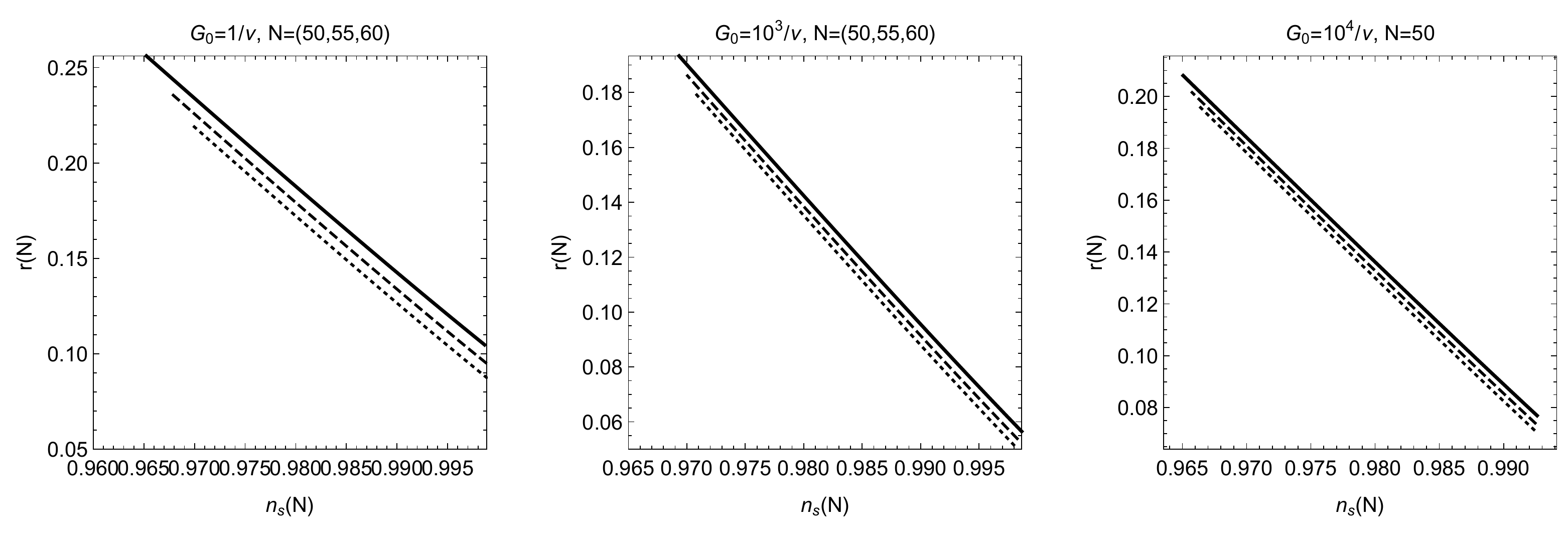}\caption{Parametric
plots for the spectral indices $\left(  n_{s},r\right)  $ in the first
approximation varying number $\nu$ and different values of $G_{0}$ and $N_{e}%
$. Left fig. is for $\left(  G_{0},N_{e}\right)  =\left(  1/\nu,\left\{
50,55,60\right\}  \right)  $, center fig. is for $\left(  G_{0},N_{e}\right)
=\left(  10^{3}/\nu,\left\{  50,55,60\right\}  \right)  $ and right fig. is
for $\left(  G_{0},N_{e}\right)  =\left(  10^{4}/\nu,\left\{
50,55,60\right\}  \right)  $. Solid lines are for $N_{e}=50,~$dashed lines are
for $N_{e}=55$ and dotted lines are for $N_{e}=60$. Left fig. is for
$\nu=\left(  1.01,1.2\right)  $, center fig. is for $\nu=\left(
2.01,2.5\right)  $, right fig. is for $\nu=\left(  2.5,3\right)  $. }%
\label{in4}%
\end{figure}

In Figs. \ref{in2}, \ref{in3} and in \ref{in4} we present the qualitative
evolution of the spectral indices according to the different values of the
free parameters. In Fig. \ref{in2} we present the contour plot for the
spectral indices $n_{s}$ and $r~$\ in the first-order approximation, in the
space of the free parameter $\left(  \nu,G_{0}\right)  $ and for $N_{e}=50$,
$N_{e}=55$ and $N_{e}=60$. we observe that as power $\nu$ increases then the
model fits the observations as $G_{0}$ increases exponentially. In Figs.
\ref{in3} and \ref{in4} we present the parametric plot in the space $\left(
n_{s},r\right)  $ for varying parameter the number of e-fold $N_{e}$, and for
specific values of the free parameters of the model $\left(  \nu,G_{0}\right)
$. In Fig. \ref{in4} we assume the that $G_{0}=\bar{G}_{0}\nu^{-1}$ and plots
are for specific values of $\nu$.

From the results of Figs. \ref{in2}, \ref{in3} and in \ref{in4} it is clear
that the new inflationary model provided by the symmetry analysis provide
values for the spectral indices according to the cosmological observations,
see Section \ref{sec2}, for the values of the Planck 2018 collaboration, Table
3 in \cite{pin1}.

\section{Conclusions}

\label{sec6}

In this work we investigated the construction of a new exact inflationary
models\ in the inflaton theory by using Lie symmetry analysis for the master
equation (\ref{ss2}). The master equation of our consideration is constructed
by the assumption the spectral index $n_{s}$ and the tensor to scalar ratio
$r$, to be related by a function~$G$. From this assumption the second-order
differential equation (\ref{ss2}) follows.

We performed a classification for the function which relates $n_{s}$ and $r$
by assuming the master equation to admit Lie symmetries. By applying the
symmetry condition we found that the master equation admits two Lie symmetries
for arbitrary function~$G$, while for $G=G_{A}$ and $G=G_{B}$ admits three Lie
symmetries and for $G=G_{C}$ the admitted Lie symmetries form the $sl\left(
3,R\right)  $ Lie algebra and the master equation is maximally symmetric. The
case $G_{C}$ was investigated in details in a previous study \cite{ref11},
hence in this work we focused on $G_{A}$ and $G_{B}$.

For $G_{A}$ and $G_{B}$, the master equation (\ref{ss2}) admits three Lie
symmetries, which form the Lie algebras $A_{3,3}$ and $A_{3,2}$ in the Patera
et al. classification \cite{pat1}, respectively. However function $G_{B}$ does
not provide a relation for the indices $n_{s}$,$~r$ as provided by the
observations, thus, we focused our analysis on the function $G_{A}$ and the
Lie algebra $A_{3,3}$.

We solved the master equation and we were able to derive the closed-form
solution for the scalar field and for all the physical variables of the
inflaton model. Furthermore, we wrote the closed-form expressions for the
spectral indices $n_{s}$,~$r$ , in the first-order approximation by using the
HSR parameters. We found that the spectral indices depend on the number of
e-fold and on two free parameters. By presenting the qualitative evolution of
the spectral indices we remark that they can fit in the cosmological observations.

\begin{figure}[ptb]
\centering\includegraphics[width=1\textwidth]{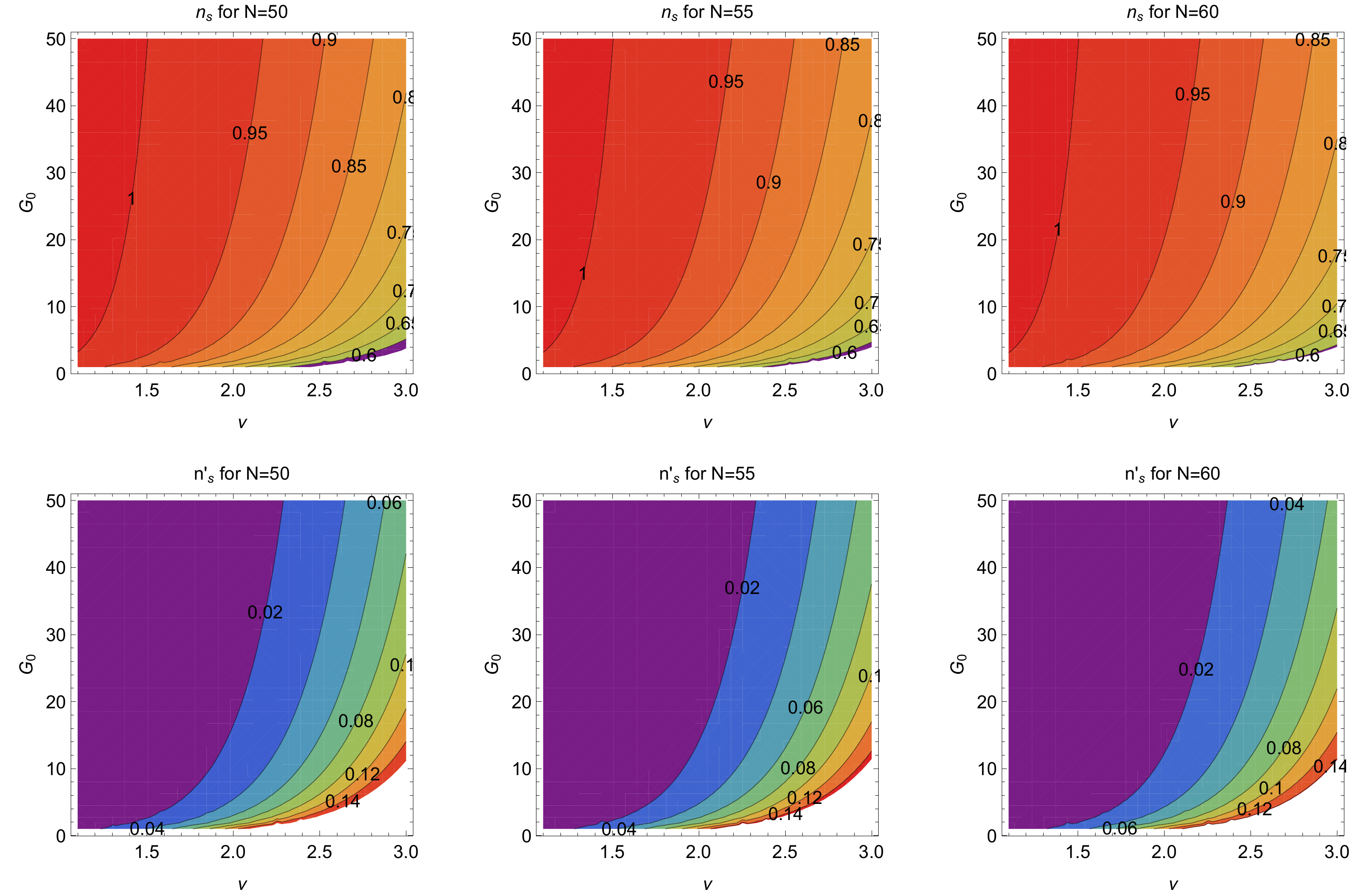}\caption{Qualitative
evolution for the spectral indices $\left(  n_{s},n_{s}^{\prime}\right)  $ in
the second approximation in the two-dimensional space for the free variables
$\left(  \nu,G_{0}\right)  $, for $N_{e}=50$ (left figs.) $N_{e}=55~$(center
figs.) and $N_{e}=60~$(right figs.) }%
\label{in5}%
\end{figure}

In Fig. \ref{in5}, we present the qualitative evolution for the spectral
indices $n_{s}$ and $n_{s}^{\prime}$ in the second approximation for different
values of the free variables $\left(  \nu,G_{0}\right)  $. We observe that
$n_{s}^{\prime}$ is always positive valued, and takes small values when
$n_{s}-1\simeq0$. According the cosmological observations, the running index
is constraint $\eta_{s}^{\prime}=-0.005\pm0.013$, therefore, small positive
values are inside the $1\sigma$ region.

We conclude that the Lie symmetry approach is a powerful method for the
derivation of exact solutions and in this study the symmetry method applied
for the derivation of a new inflationary exact solution. A natural question
which raised is if this approach can be applied in $f\left(  R\right)
$-theory or in scalar-tensor gravity \cite{ff3}. These three theories are
related under conformal transformations. The analytical solution determined in
\cite{ns1} for the scalar field can be easily used to write the corresponding
analytic solutions for the conformal equivalent theories, and extend the
present symmetry analysis in the conformal frame. However, such an analysis
overpass the scopus of the present work and will be discussed elsewhere. 

In a future study we plan to investigate further this scalar field solution
model and we want to use this model as dark energy candidate and investigate
if it solves the $H_{0}$-tension \cite{h01}.

\begin{acknowledgments}
This work is based on the research supported in part by the National Research
Foundation of South Africa (Grant Number 131604).
\end{acknowledgments}

\end{document}